# Electronic origin of melting T–P curves of alkali metals with negative slope and minimum

Degtyareva V.F.

*Institute of Solid State Physics, Russian Academy of Sciences, Chernogolovka, Russia*

*degtyar@issp.ac.ru*

Group I elements – alkali metals Li, Na, K, Rb and Cs – are examples of simple metals with one *s* electron in the valence band. Under pressure these elements display unusually complex structural behaviour transforming from close-packed to low symmetry open-packed structures. Unexpectedly complex form was found for melting curves of alkalis under compression with an initial increase in the melting temperature in accordance with Lindemann criterion and a further decrease to a very low melting point. To understand complex and low symmetry crystal structures in compressed alkalis, a transformation of the electron energy levels was suggested which involves an overlap between the valence band and outer core electrons. Within the model of the Fermi sphere – Brillouin zone interaction one can understand the complex melting curve of alkalis.

## 1. Introduction

Group I elements – alkali metals Li, Na, K, Rb and Cs – are examples of *simple* metals with one *s* electron in the valence band. Experimentally it was found that under pressure these elements display unusually complex structural behaviour transforming from close-packed structures such as body-centered cubic (bcc) and face-centered cubic (fcc) to low symmetry open-packed structures (see review [1] and references therein). An unexpectedly complex behaviour was found for T – P melting curves of alkalis under compression with an initial increase in the melting temperature in accordance with Lindemann criterion and a further decrease to very low melting point. The lowest point of melting was found to be 300K at 120 GPa for sodium and 200K at 50 GPa for lithium [2,3]. Heavier alkalis have also non-monotonic behaviour of the melting curve on T – P diagrams.

Complexity of T – P phase diagrams of alkali elements attracts attention of theoretical considerations. To mention some aspects of the theoretical approaches that are related to the focus of the present paper, the Fermi surface – Brillouin zone interaction was found to be a "dominant contribution" to crystal structure of s-band elements including alkali metals [4]. Essential deformation of the Fermi surface from the initially spherical form was calculated for lithium [5]. The negative melting slope of sodium under pressure was reproduced with calculations as driven by electronic transitions in the liquid state of Na [6]. The melting curve of dense sodium was simulated by assuming effects of inner-core $2p$ states [7].

To understand complex and low symmetry structures in compressed alkalis a transformation of the electron energy levels was suggested which involves an overlap between the valence band and outer core electrons [8,9]. The structures Na-$oP8$ above 117 GPa and K-$oP8$ above 54 GPa are similar to the intermetallic compound AuGa at ambient pressure with the 2 valence electrons per atom. From this similarity it was suggested that Na and K became divalent metals under strong compression. Within the model of the Fermi sphere – Brillouin zone interaction one can understand the complex melting curve of alkalis.

## 2. Theoretical background and method of analysis

The crystal structure of metallic phases is defined by two main energy contributions: electrostatic (Ewald) energy and the electron band structure term. The latter usually favours the formation of superlattices and distorted structures. The energy of valence electrons is decreased due to a formation of Brillouin planes with a wave vector q near the Fermi level $k_F$ and opening of the energy pseudogap on these planes if $q_{hkl} \approx 2k_F$. Within a nearly free-electron model the Fermi sphere radius is defined as $k_F = (3\pi^2 z/V)^{1/3}$, where z is the number of valence electrons per atom and V is the atomic volume. This effect, known as the Hume-Rothery mechanism (or electron concentration rule), was applied to account for the formation and stability of the intermetallic phases in binary simple metal systems like Cu-Zn, and then extended and widely used to explain the stability of complex phases in various systems, from elemental metals to intermetallics [10].

The stability of high-pressure phases in alkalis is analyzed using a computer program BRIZ [11] that has been developed to construct Brillouin zones or extended Brillouin-Jones zones (BZ) and to inscribe a Fermi sphere (FS) with the free-electron radius $k_F$. The resulting BZ polyhedron consists of numerous planes with relatively strong diffraction factor and accommodates well the FS. The volume of BZ's and Fermi spheres can be calculated within this program.

The liquid state of metals can be considered within the FS – BZ model assuming that the first strong structure factors correspond to a nearly-spherical BZ [12]. The interaction of this BZ with the Fermi level at $k_F$ leads to the formation of the pseudo-gap and a decrease of the electronic energy. The well known example of significant increase of stability for liquid state is displayed by several binary alloys with the eutectic type of phase diagrams such as Au – Ge and Au – Si. These alloys have the melting temperature at around $360^{o}$C whereas the pure Au melts at $1064^{o}$C. The composition of eutectic alloys is 28 at% Ge and ~20 at% Si corresponding to z = 1.84 or 1.6 electron/atom, respectively.

Mercury – the only elemental metal that exists in the liquid state at ambient conditions – has two electrons in the valence band. The melting point of Hg is $–38.7^{o}$C. This and the above examples show the increased stability of the liquid state in the metals with 1.6 – 2 valence electrons. The structure factor for liquid metals is deformed by appearance of a shoulder on the high-Q side as was observed for Hg [13]. Stability of the liquid state in Hg was analysed by appearance of a deep minimum of density of states near the Fermi level [14].

## 3. Evaluation of experimental data

### 3.1. Flat and negative slope of melting curves of alkalis

Unusual behaviour of the T-P curves of alkalis may result from a strong increase of band structure energy where the nearly spherical Fermi surface of bcc-alkalis develop necks attracted to BZ planes, as was shown for *bcc*-Li at 8 GPa [5]. However, this attraction is reciprocal and leads to contraction of BZ planes and to expansion of the unit cell in the real space. FS-BZ interactions exist also for liquids that have spherical BZ and non-directional attraction to FS. Therefore FS-BZ interactions produce more remarkable effects for solid bcc-state and the liquid state becomes denser than crystal at some pressure. Consequently, the melting curve display a negative slope as is seen on Figure 1 for sodium around 30 – 40 GPa. At these pressures the atomic local order for bcc-Na and liquid-Na is similar and there is no electronic transfer because of a lack of *d*-states in the valence shell in contrast to Cs, as discussed in [6]. At higher pressure above 65 GPa the electronic transition in liquid-Na occurs at lower pressures than for solid-Na [6].

### 3.2. Deep minimum of melting curves of alkalis

We suggest that the decrease of the melting temperatures is related to an overlap of the valence electron band and core electrons as was considered for the structures Na-*oP*8 above 117 GPa and K-*oP*8 above 54 GPa where it was assumed that Na and K become divalent metals at such compression [8,9]. It is known that the liquid reacts to the external influence more easily than the solid. Similar electron transfer may occur in liquid state and even at lower pressure than for solid state.

Experimental diffraction data on liquid Na near the minimum of T-P melting curve (from [15]) display structure factor peaks with appearance of a shoulder similar to that of liquid Hg as shown in Figure 2. Therefore it is reasonable to assume the valence electron number for Na at this pressures equals two (like Hg), whereas for solid Na (*fcc* and *cI*16 phases) at this pressure the valence electron number equals to one.

Diffraction data for Rb in the liquid state (from [16]) display significant changes under compression as shown in Figure 3. The first structure factor undergoes broadening with appearance of a shoulder at the high-Q side becoming similar to that of liquid Hg. These changes can be attributed to the suggestion of the increase of the number of electrons in the valence band at the expense of the outer core electrons.

Similar effects have been observed experimentally for liquid Cs as shown on Figure 4 (after data from [17]). In the heaviest alkali metal Cs the electron transfer starts at significantly lower pressure just above 4 GPa, as can be suggested by analyzing the broadening of the first diffraction peak. This event of diffraction changes in the liquid state under compression was observed also for potassium [18]. Thus, the structural and electronic transfer is common for all dense liquid alkalis.

## 4. Conclusion

Long sequences of crystal structures for alkali metals under compression with the transformations from the close-packed structures to the complex open-packed structures and non-monotonic T - P curves raise the following question: what are the physical reasons behind these behaviours. One of reason that we suggest in this paper is the Fermi sphere – Brillouin zone interactions that increases under pressure.

The first stage of T – P melting curves with an appearance of a maximum that is followed by a negative slope can be explained due to difference in the attraction of FS to BZ that occurred in crystal and liquid states and causes the liquid to become more compressible than solid.

The next stage of T - P melting curves with deep minimum and unusual increase of stability of liquid phase can be understood by assuming the increase of the valence electron numbers due to the overlap with the core electrons under strong compression. Non-simple behaviour in melting of alkali metals under high pressure is connected to the essential changes of the electron state in the valence band.

**Acknowledgements**

The author gratefully acknowledges Olga Degtyareva for her valuable discussion and comments. This work was supported by the Program of the Russian Academy of Sciences "The Matter under High Pressure".

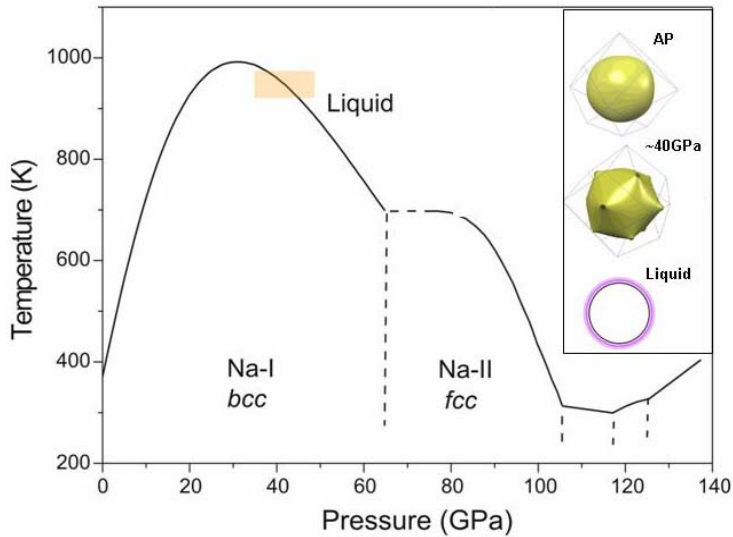

Figure 1. The T-P melting curve for Na (after data from [2]) with the negative slope above 30 GPa. In the inset, Fermi surface – Brillouin zone configurations are given for Na-*bcc* at ambient pressure (AP) and at ~40 GPa in comparison with the nearly spherical pseudo-BZ for the liquid Na.

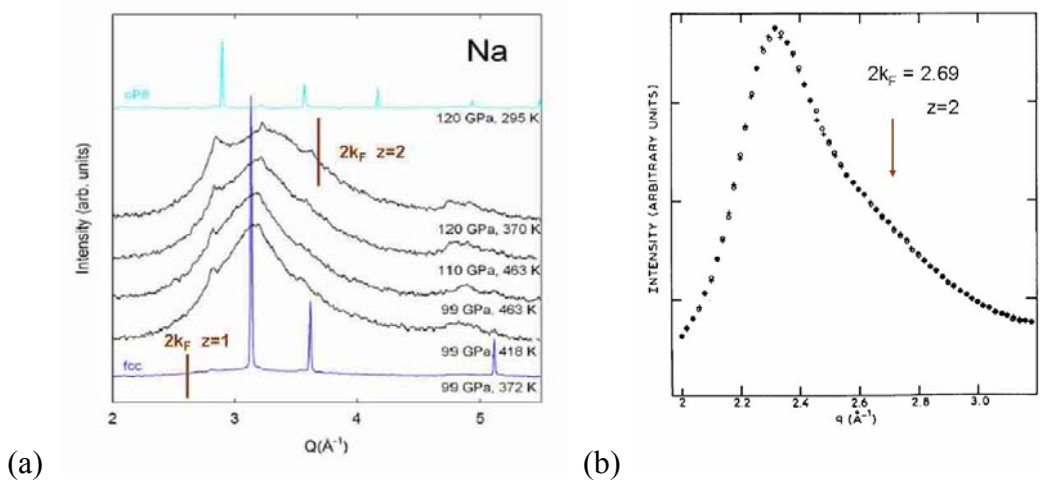

(a)                                                    (b)

Figure 2. (a) Diffraction data for Na in solid and liquid state (after data from ESRF report [15]). Evaluation of 2kF positions is given by accepting the valence electron number z = 1 for Na-fcc and z = 2 for Na-liquid at 120 GPa. (b) Diffraction data for liquid Hg at ambient condition (after data from [13]) with indication of 2kF positions for z = 2.

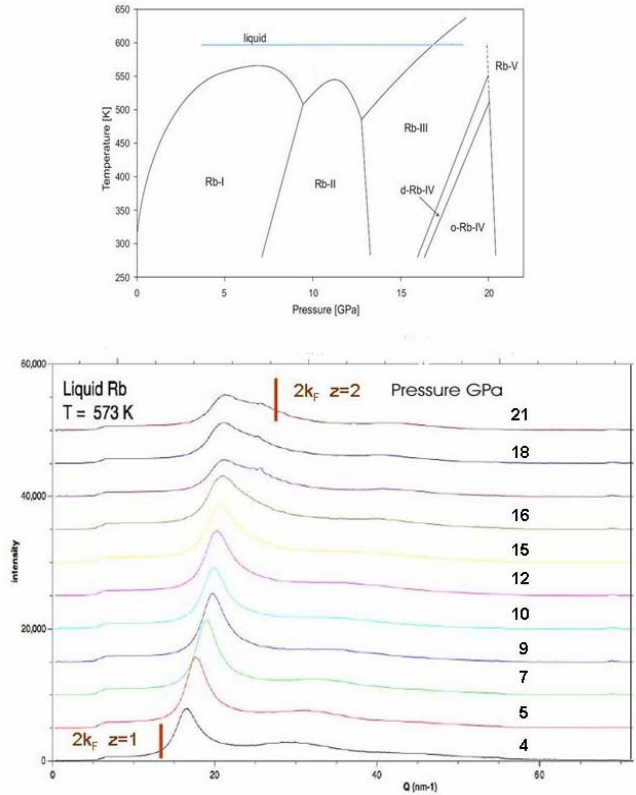

Figure 3. Diffraction data for Rb in the liquid state (after data from ESRF report [16]). The trace of measured points is indicated on the phase T-P diagram (upper plot). Evaluation of $2k_F$ positions is given by accepting the valence electron number $z = 1$ for 4 GPa and $z = 2$ for 21 GPa.

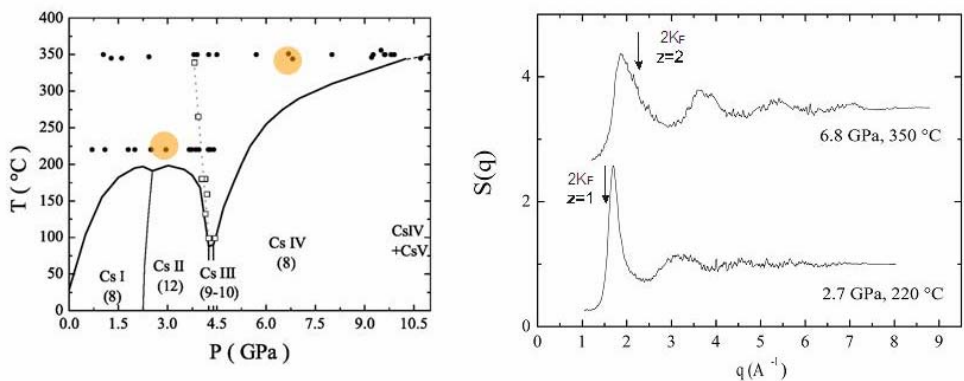

Figure 4. The T-P phase diagram for Cs on the left panel with the marked experimental points (red) for diffraction data for Cs in the liquid state on the right panel (after data from [17]). Evaluation of $2k_F$ positions is given by accepting the valence electron number $z = 1$ for 2.7 GPa and $z = 2$ for 6.8 GPa.